\documentclass{article}
\usepackage{epsfig}
\usepackage{calc,xcolor}

\newcommand{\keyw}[1]{{\bf #1}}

\newcommand{\dahntab}[1]{
  \newbox\mybok%
  \setbox\mybok=\hbox{\vbox{
      \begin{tabbing}
        #1
      \end{tabbing}%
    }}

  \newdimen\bokwidth%
  \bokwidth=\wd\mybok%
  \newdimen\myl%
  \myl=\textwidth%
  \divide\myl by 2%
  \divide\bokwidth by -2%
  \advance\myl by\bokwidth%
  \vrule width\myl height 0pt depth 0pt%
  \usebox\mybok%
}

\newtheorem{lemma}{Proposition}
\newtheorem{definition}{Definition}

\newtheorem{theorem}{Theorem}
\newtheorem{remark}{Remark}
\def\beginproof{\noindent{\bf Proof.}\quad}
\def\endproof{$\diamondsuit$}

\definecolor{grey}{gray}{0.85}

\def\goesto{\rightarrow}

\begin{document}
\setcounter{page}{1}

\title{On the dynamics of Social Balance on general networks (with an application to XOR-SAT)}
\author{Gabriel Istrate \\eAustria Research Institute, \\ Bd. V. P\^{a}rvan 4, cam. 045B, \\Timi\c{s}oara RO-300223, Romania.\\ Email: {\tt gabrielistrate@acm.org}}

\maketitle

\begin{abstract}
We study nondeterministic and probabilistic versions of a discrete
dynamical system (due to T. Antal, P. L. Krapivsky, and S. Redner
\cite{redner-balance}) inspired by Heider's social balance theory.
We investigate the convergence time of this dynamics on several
classes of graphs.  Our contributions include:
\begin{enumerate}
\item We point out the connection between the triad
dynamics and a generalization of {\em annihilating walks}
to hypergraphs. In particular, this connection allows us
to completely characterize the recurrent states in graphs
where each edge belongs to at most two triangles.
\item We also solve the case of hypergraphs that do not contain edges consisting of one or two vertices.
 \item We show that on the so-called ``triadic
cycle''graph, the convergence time is linear.
\item We obtain a cubic upper bound on the convergence
time on 2-regular triadic simplexes $G$. This bound can be further
improved to a quantity that depends on the {\em Cheeger constant} of
$G$. In particular this provides some rigorous counterparts to
experimental observations in \cite{triad-lattice}.
\end{enumerate}

We also point out an application to the analysis of the random walk algorithm on certain instances of the 3-XOR-SAT problem.
\end{abstract}



\pagestyle{myheadings}
\thispagestyle{plain}
\markboth{G. ISTRATE}{DYNAMICS OF SOCIAL BALANCE}

\maketitle

\section{Introduction}

Discrete Dynamical Systems, both synchronous (a.k.a. cellular
automata \cite{ilachinski-ca}) and asynchronous (e.g. sequential
dynamical systems \cite{sds-book}) provide a rich family of models,
whose computational properties (e.g. computational universality, the
complexity of prediction, or properties of the phase space, such as
the existence of Garden of Eden states) have been actively
investigated. Such models underline some of the most celebrated
examples in the social sciences (e.g. the celebrated Schelling
Segregation Model) and have, unsurprisingly, been proposed (see e.g.
\cite{hegselmann-flache-ca-jasss}, \cite{sds-social}) as the basis
for modeling social dynamics.

In this paper we are concerned with the properties of a particular
dynamics, due to T. Antal, P. L. Krapivsky and S. Redner
\cite{redner-balance}. The dynamics (referred in the sequel as {\em
the triadic dynamics}) is inspired by Heider's {\em balance theory}
\cite{heider-book}, a well-established subject in social psychology,
and one of the first theories in the social sciences to benefit from
concepts and methods from graph theory \cite{cartwright-harary}.

Heider's balance theory aimed to describe the ``equilibrium''
properties of the interpersonal relations, and did not include a
dynamical component. Recently, however, dynamical models related to
the one we study, have become popular in the statistical physics
\cite{heider-continuous,heider-simulate} and social simulation
literature  \cite{hummon-doreian-sn,heider-jasss}.

 The triadic dynamics (described precisely in Section
2 below) is parametrized by a constant $p\in [0,1]$. In
\cite{redner-balance} the authors investigated the dynamics on the
complete graph $K_{n}$ and displayed a double phase transition in
the convergence time with respect to control parameter $p$. The
triadic dynamics was further investigated by means of computer
simulations in \cite{triad-lattice}. The graph topology in this
paper is a finite section of the triangular lattice. A phase
transition is also displayed around a critical value $p_{C}\sim
0.4625$. In this case the convergence time is polynomial for
$p<p_{C}$ and logarithmic for $p>p_{C}$.

We will be concerned with extending the study of the triadic
dynamics to general graphs. To do so, we have to solve a  problem
that has appeared many times in the study of dynamical systems, that
of {\em state reachability}: when is a given system configuration
$s_{2}$ reachable from a given initial state $s_{1}$ by a finite
sequence of moves ? In general (e.g. \cite{sds-complexity, BH+03})
this type of problem is computationally intractable even for very
simple dynamical systems. In contrast our main result shows that for
a large class of graph topologies questions such as reachability are
computationally tractable for our dynamics. We complement this
result with a study of the convergence time of our dynamics in the
probabilistic setting similar to that of \cite{redner-balance}.

A completely different reason for our interest in the triadic
dynamics is its somewhat unexpected relation \cite{triad-xorsat} to
{\em 3-XOR-SAT}, a combinatorial problem that was investigated in
Theoretical Computer Science \cite{dubois-mandler-3xor} and
Statistical Physics \cite{zecchina:kxorsat}. In particular, using
the methods we develop for the analysis of the dynamics we will
upper bound the expected convergence time of a simple local search
algorithm, called {\em RandomWalk}, previously analyzed for random
3-XOR-SAT in \cite{semerjian-monasson-xor,barthel-hartman-weigt}.
Unlike these papers, we will analyze the convergence time of the
algorithm on individual formulas, and bound the convergence time
(Theorem~\ref{time-rw} below) using quantities that depend on the
structure of the input formula.

\section{Preliminaries} The following is a formal definition of the dynamics we will be investigating in this paper:

\begin{definition} \label{nondet}
 {\bf Nondetermininstic Triadic Dynamics}. We start with a
graph $G=(V,E)$ whose edges are labeled $\pm 1$. A
triangle $T$ is $G$ is called {\em balanced} if the
product of the labels of its edges is equal to 1. At any
step $t$, for any imbalanced triangle $T$ we are allowed
to change the sign of an arbitrary edge of $T$ (thus
making $T$ balanced). The move might, however, make other
triangles unbalanced.
\end{definition}
\begin{definition}
 {\bf Probabilistic Triadic Dynamics}. A probabilistic
version of the dynamics in Definition~\ref{nondet},
parametrized by a real number $p\in (0,1)$, is specified
\cite{redner-balance} as follows: While there exists an
imbalanced triangle, first choose uniformly at random an
imbalanced triangle $T$. If the triangle $T$ has a single
negative edge $e$:
\begin{itemize}
\item With probability $p$ turn $e$ to positive.
\item With probability $1-p$ turn one of the other two edges of $T$ to negative.
\end{itemize}
otherwise ($T$ has three imbalanced edges) change the label of a random edge of $T$.
\end{definition}

\begin{definition}

 A {\em triadic simplicial complex} is a graph $G=(V,E)$ such that all edges $e\in E$ are part of some triangle of $G$.
\end{definition}

\begin{definition}
 The {\em triadic dual } of graph $G$ is an undirected
hypergraph with self-loops
$T_{3}(G)=(\overline{V}_{3},\overline{E}_{3})$ defined as
follows: $\overline{V}_{3}$ is the set of triangles of
$G$. Hyperedges in $T_{3}(G)$ correspond to edges in $G$
and connect all vertices $v\in \overline{V}_{3}$
containing a given edge. In particular
we add a self-loop to vertex $v\in \overline{V}_{3}$ if
$v$ belongs to an unique triangle of $G$. We may even add
two self-loops to the same vertex $T$ if two of its edges belong only to triangle $T$.
\end{definition}

We will also require that the triadic dual of graph $G$ be connected. If this was not the case, the dynamics would decompose on independent dynamics on the connected components.

\begin{definition}
 The {\em triadic cycle $TC_{n}$} is the graph $G$
consisting of $n$ triangles chained together. Edges $AB$
and $CD$ of the extreme triangles in the chain are
``glued'' by identifying $A\equiv C$, $B\equiv D$.
\end{definition}

The construction is illustrated in
Figure 1, where the triadic cycle graph
with sixteen triangles is displayed, together with its
triadic dual (in this graph circles represent self-loops).

\begin{figure}[ht]\label{triadic-cycle}
\hfill
  \begin{minipage}[t]{.45\textwidth}
\begin{center}
\epsfig{file=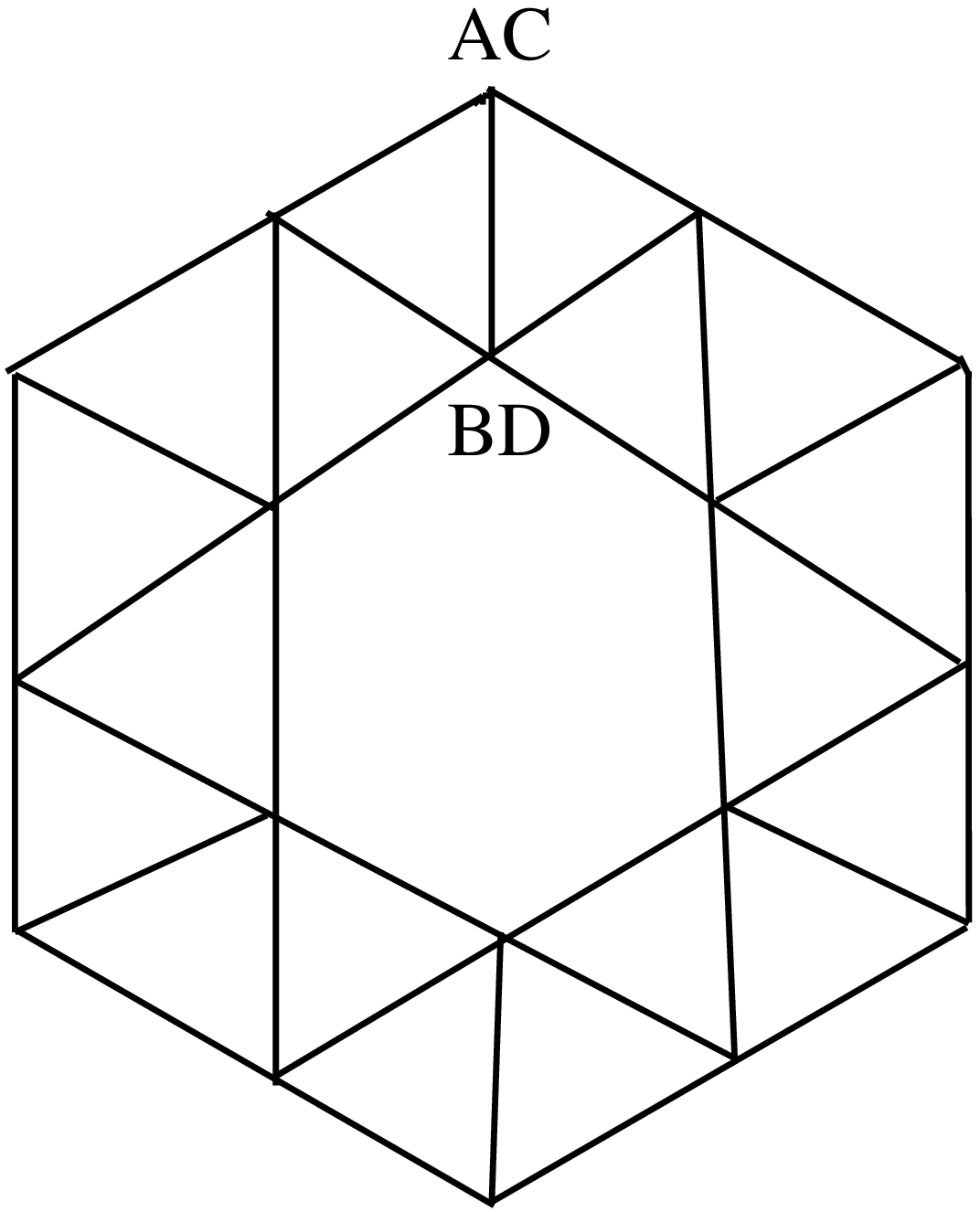,width=6cm}
\end{center}
  \end{minipage}
  \hfill
  \begin{minipage}[t]{.45\textwidth}
    \begin{center}
\epsfig{file=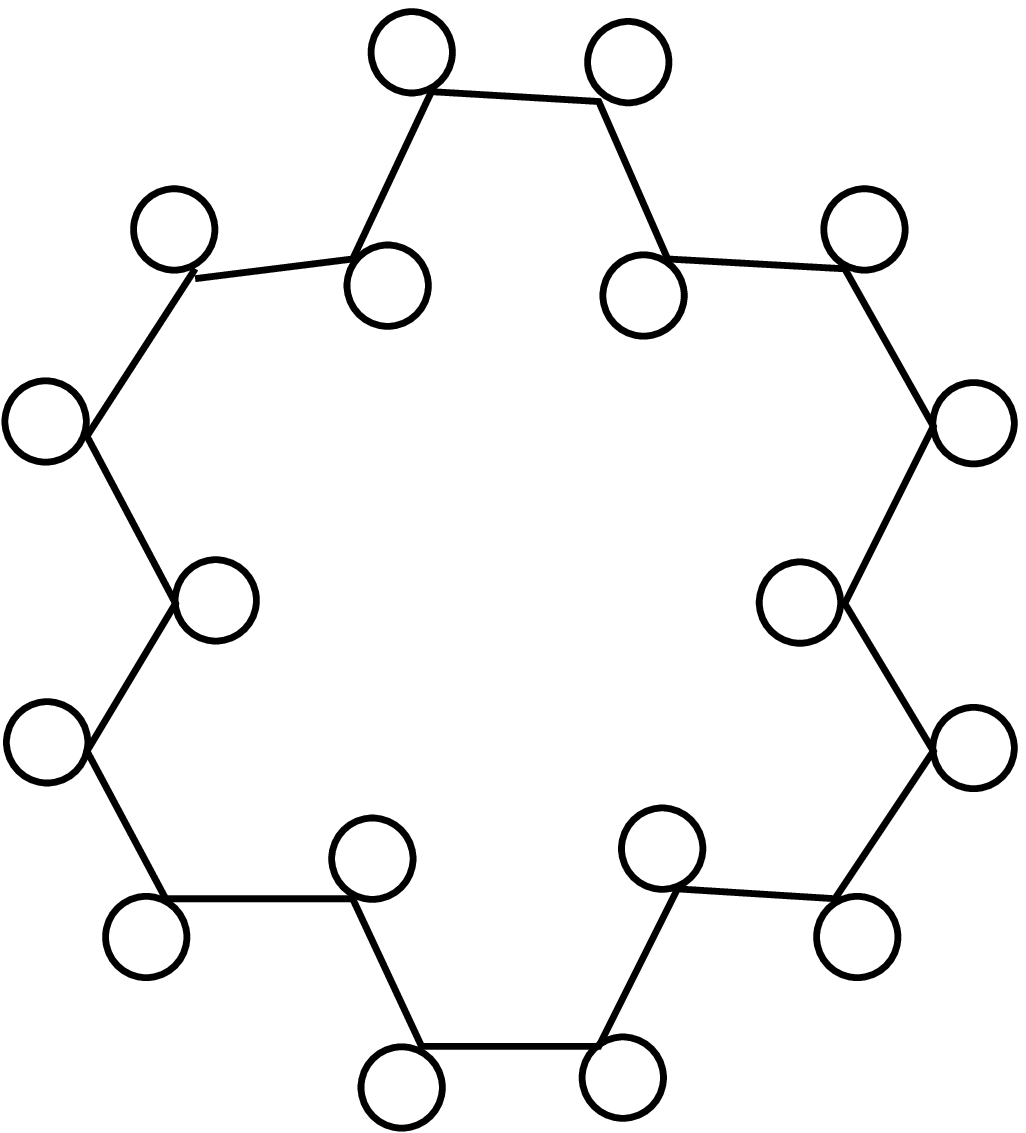,width=6cm}
\end{center}
  \end{minipage}
\caption{(a). The triadic cycle $TC_{18}$. (b). Its triadic dual}
\end{figure}

\begin{definition} Let $G=(V,E)$ be a graph. For $S\subseteq V$ denote by
$vol(S)=\sum_{x\in S} d(x)$ and $E(S,\overline{S})$ the set of edges connecting a
 node in $S$ to one in $\overline{S}$.

The {\em Cheeger time of graph $G$} is
\[
 \tau_{c}(G)=\sup_{A\subseteq V}\frac{\pi[A]\cdot \pi[\overline{A}]}{\sum_{i\in A}\sum_{j\in \overline{A}}\pi[i]p_{i,j}},
\]

where $\pi[A]=\frac{|A|}{n}$ and, for $i,j\in V(G)$, we define
$p_{i,j}=1/deg(i)$ iff $i$ and $j$ are adjacent, $0$ otherwise.

\end{definition}

\begin{remark}
 When $G$ is $r$-regular $\tau_{c}(G)=\sup_{A\subseteq V}\frac{|A|\cdot |\overline{A}|/n^2}{r/n\cdot |E(A,\overline{A})|}=\sup_{A\subseteq V}\frac{|A|\cdot |\overline{A}|}{rn\cdot |E(A,\overline{A})|}$.
\end{remark}

Finally, we need the following concept from  \cite{aldous-meeting}:

\begin{definition} Let $(X_{n}),(Y_{n})$ be independent random walks on $G$
 The worst case meeting time of $G$ is
defined as:
\[
 \tau_{M}(G)=\max_{i,j\in V(G)}E[T_{M}|X_{0}=i,Y_{0}=j],
\]
where $T_{M}=\min\{k: X_{k}=Y_{k}\}$.

\end{definition}

\section{Connection with annihilating walks}

The triad dynamics can be mapped to a generalization of {\em annihilating walks}.

\begin{definition} \label{aw} An {\em annihilating walk
(AW) on graph
$G$}  is a nondeterministic dynamical system described as
follows: Start with a ball at some of the vertices of the
random graph. At each step we move one ball to a neighbor
of its current vertex. When two balls meet at a vertex $v$
they annihilate each other, and are subsequently
eliminated from graph $G$.
\end{definition}

The stochastic version of annihilating walks (called {\em
annihilating random walks} \cite{erdos-ney-arw}) have been
studied in the  interacting particle systems literature (
e.g. \cite{griffeath-ips}). A few recent results
\cite{donnelly-welsh-interacting},\cite{aldous-fill-book}
deal with  interacting particle systems on a finite graph
as well.

We will also need the following:

\begin{definition} A {\em coalescing random walk (CRW) on graph $G$} is a stochastic process described as follows:
\begin{enumerate}
 \item Start with a ball at some of the vertices of the random graph.
\item Each ball performs a random walk on $G$.
\item When two or more balls meet at a vertex $v$ they coalesce, performing
a common random walk on $G$.
\end{enumerate}
\end{definition}
\begin{definition}
For an annihilating random walk
we denote by $T_{ARW}(G)$ the random variable defined as
the number of steps until the number of balls on graph $G$
becomes zero (one if the initial number of balls was odd).
For a coalescing random walk we denote by $T_{CRW}(G)$ the
random variable defined as the number of steps until all
the balls coalesce.
\end{definition}

To recover the connection between annihilating walks and
triad dynamics, we have to define a generalization of
annihilating walks to hypergraphs.

\begin{definition} \label{cb} The {\em hyperedge switching
process (HS) on a hypergraph
$G$}  is a nondeterministic dynamical system described as
follows: Start with a ball at some of the vertices of the
hypergraph. At each step we choose a hyperedge
$C$ in $G$ that contains at least a ball. Any self-loop on
a  vertex that contains a ball counts as one of these such
cliques. We then put balls
on all empty nodes of $C$ and remove the balls from the
nonempty nodes.
\end{definition}

\begin{lemma}
 The nondeterministic triad dynamics on graph $G$ and initial configuration
 $s$ corresponds to the hyperedge switching process on graph $T_{3}(G)$ to the initial
 configuration $\overline{s}$ defined by putting a ball to all vertices $v\in T_{3}(G)$
 corresponding to unbalanced triangles
 in $G$.
\end{lemma}
\beginproof
This is immediate by duality. Indeed, choosing an edge $e$ to flip
(in the original graph) corresponds to chosing the corresponding
hyperedge $l_{e}$ (in its dual hypergraph). The choice is only
possible if there exists at least one imbalanced triangle containing
$e$, that is (by duality) at least one of the vertices of $l_{e}$
contains a particle. Flipping the sign of edge $e$ only affects
triangles containing it (i.e. vertices of $l_{e}$ in the triadic
dual), and has the effect of making imbalanced triangles balanced
and viceversa. This corresponds plainly to removing balls from the
occupied vertices of $l_{e}$ and adding balls to the free vertices.
\endproof

\section{Recurrent states for nondeterministic triad
dynamics}

\begin{definition}
A state $s$ is specified by giving a $\pm 1$ label to each
edge of graph $G$. A state is {\em recurrent} if for any
infinite path $\pi=(\pi_{0},\ldots, \pi_{n}\ldots)$ of the
system $s$ is reachable from $\pi_{n}$.
\end{definition}

It is fairly easy to determine recurrent states if every
edge is part of at most two triangles (that is, if the
triadic dual is a graph with loops):

\begin{theorem}\label{recurrent}
 Let $G$ be a graph such that (i) the triangle graph of $G$ is connected, and (ii) every edge of $G$ is part of at most two triangles, and let $s_{0}:E(G)\goesto \{\pm 1\}$ be an initial state. Then:

\begin{enumerate}
 \item If $G$ contains an edge that is part of only one triangle, then the recurrent states for the triad dynamics are exactly the {\em completely balanced states} reachable from $s_{0}$.
\item If every edge in $G$ is part of exactly two triangles and the number of imbalanced triangles in $s_{0}$ is even then the recurrent states for the triad dynamics are exactly the {\em completely balanced states} reachable from $s_{0}$.
\item If every edge in $G$ is part of exactly two triangles and the number of imbalanced triangles in $s_{0}$ is odd, then the recurrent states for the triad dynamics are the states containing {\em exactly} one imbalanced triangle.
\end{enumerate}
\end{theorem}

\beginproof

Define function $P$ that maps each state $s$ to the number of imbalanced triangles in state $s$.
\begin{lemma}\label{dec}
 Function $P$ is nondecreasing. That is, if $s\goesto t$ is a legal transition of the dynamics, then $P(s)\geq P(t)$.
\end{lemma}
\beginproof
Consider the edge $e\in G$ chosen by the dynamics. There are three cases:

\begin{enumerate}
\item There is only one (imbalanced) triangle $T$ that contains $e$. Then, by flipping the sign of edge $e$ triangle $T$ becomes balanced and no other triangle is affected. The number of imbalanced triangles goes down by one.
 \item There are two triangles $T_{1},T_{2}$ containing edge $e$ (Figure~\ref{two-triangles}), both imbalanced. Then, by flipping the sign of edge $e$ triangles $T_{1},T_{2}$ become balanced and the balancedness of no other triangle is affected. Hence the number of imbalanced triangles goes down by two.
\item There are two triangles $T_{1},T_{2}$ containing edge $e$, one imbalanced (say it is $T_{1}$) and one ($T_{2}$) balanced. Then, by flipping the sign of edge $e$ triangle $T_{1}$ becomes balanced, while $T_{2}$ becomes imbalanced. The balancedness of no other triangle is affected. Thus the number of unbalanced triangles stays the same.
\end{enumerate}
\endproof

Consider now a recurrent state $t$. It follows that there is no state $w$ reachable
from $t$ with $P(w)<P(t)$.

To prove 1 note that from any state containing some imbalanced triangle we can
reach a state $w$ with one less balanced triangle by ``propagating the imbalance''
towards the triangle $T$ with an edge $e$ occurring only in $T$. In this process the
number of imbalanced triangles stays the same. By then
changing the label of $e$ we decrease the number of imbalanced triangles.

The arguments for points 2 and 3 are similar, noting that the number of imbalanced
triangles stays the same or decreases by exactly 2.
\endproof

We would like to extend our result to general graphs $G$.
However, this is an open problem so far. The reason is that for general graphs $G$ the triadic dual is no longer a graph, but a hypergraph and the potential function is no longer nonincreasing.
In the sequel we provide a partial result.

\subsection{Recurrent configurations for the nondeterministic hyperedge switching process}

In this section we study the hyperedge switching process on
hypergraphs $H$ whose hyperedges all contain more than two vertices.
We will refer to this condition as {\em $H$ does not contain graph
edges}.

\begin{definition}
For every pair of boolean configurations
$w_{1},w_{2}:V(H)\rightarrow {\bf Z}_{2}$ on hypergraph $H$ we
define a system of boolean linear equations $H(w_{1},w_{2})$ as
follows: Define, for each hyperedge $e$ a variable $z_{e}$ with
values in ${\bf Z}_{2}$. For any vertex $v\in V(H)$ we define the
equation
\begin{equation}\label{system}
 \sum_{v\in e}z_{e}=w_{2}(v)-w_{1}(v).
\end{equation}
In equation~\ref{system} the difference on the right-hand side is
taken in ${\bf Z}_{2}$; also, we allow empty sums on the left side.
System $H(w_{1},w_{2})$ simply consists of all equation
(\ref{system}), for all $v\in V(H)$.
\end{definition}

\begin{definition}
If $x$ is a state on $H$ and $l$ is an edge of $H$, define
\begin{equation}
 x^{(l)}(v)=\left\{\begin{array}{cc}
               1+x(v),&\mbox { if }v\in l, \\
           x(v),&\mbox{ otherwise.}\\
              \end{array}
       \right.
\end{equation}
\end{definition}

In this section we prove the following:

\begin{theorem}\label{trees}
 Let $H$ be a connected hypergraph with no graph edges, let $w_{1}$
 be an initial configuration that is not identical to the "all zeros" configuration {\bf 0},
 and let $w_{2}$ be a final configuration.

There is a polynomial time algorithm to test whether  $w_{2}$ is a recurrent state for the nondeterministic hyperedge process on $H$ with starting state $w_{1}$.
\end{theorem}
\beginproof

\begin{lemma} \label{if}
 If state $w_{2}$ is reachable from $w_{1}$ then the system of equations $H(w_{1},w_{2})$ has a solution in ${\bf Z}_{2}$.
\end{lemma}
\beginproof
Let $P$ be a path from $w_{1}$ to $w_{2}$ and let $z_{e}$ be the number of times edge $e$ is used on path $P$ (mod 2). Then $(z_{e})_{e\in E}$ is a solution of system~(\ref{system}). Indeed, element $w(v)$ (viewed modulo 2) flips its value anytime an edge containing $v$ is scheduled.
\endproof

Lemma~\ref{if} gives a necessary condition for reachability in  Theorem~\ref{trees}. The complete characterization of recurrent states is a consequence of the following two lemmas:

\begin{lemma}\label{onlyif} Let $H$ be a hypergraph with no graph edges (not  necessarily connected). Assume that $w_{1},w_{2}\in \{0,1\}^ {V(H)}$ are configurations such that for no connected component $C$ of $H$, $w_{1}|_{C}\equiv {\bf 0}$, $w_{2}|_{C}\not \equiv {\bf 0}$. Then state $w_{2}$ is reachable from $w_{1}$ {\bf if and only if} system of equations  $H(w_{1},w_{2})$ has a solution in ${\bf Z}_{2}$. 
\end{lemma}
\beginproof
We prove the result by induction on the number of edges in hypergraph $H$.
\begin{enumerate}
 \item {\bf Case $m=1$:} Suppose system $H(w_{1},w_{2})$ has a solution. Since $H$ contains a single edge $e$, $w_{2}(v)=w_{1}(v)$ for all vertices $v\not \in e$ (otherwise the system would contain equation $0=1$). Also, for all $v\in e$ the quantity $w_{2}(v)-w_{1}(v):=\lambda $ does not depend on $v$ (otherwise we would have two contradicting equations). There are two cases: $\lambda = 0$, in which case $w_{1}=w_{2}$ and path $P$ can be taken as the empty path, and the case $\lambda =1$. In this case it follows that edge $e$ contains at least one vertex $v$ with $w_{1}(v)=1$. Indeed, if this was not true then $w_{1}|_{e}\equiv {\bf 0}$, and $w_{2}|_{e}\equiv {\bf 1}$, contradicting the restriction from the hypothesis. One can then set $P=\{e\}$ and obtain a path from $w_{1}$ to $w_{2}$.
\item {\bf Case $m\geq 2$:}

Assume Lemma~\ref{onlyif} is true for all  hypergraphs with less than $m$ edges, and consider an arbitrary  hypergraph $H$ with $m$ edges. Without loss of generality we may assume that $H$ is connected, otherwise we find paths separately on every connected component.  There are two cases:
\begin{enumerate}
 \item {\bf  For some $l\in E$ system $H(w_{1},w_{2})$ has a solution with $z_{l}=0$}.

Then the system $U(w_{1},w_{2})$, corresponding to hypergraph $U=H\setminus \{l\}$ is solvable; indeed every solution of $H(w_{1},w_{2})$ with $z_{l}=0$ is also a solution of $U(w_{1},w_{2})$.

Hypergraph $U$ has four types of connected components:
\begin{enumerate}
\item Connected components $P$ of $U$ such that $w_{1}|_{P\setminus\{l\}}\neq {\bf 0}$.
\item Connected components $Q$ of $U$ such that $w_{1}|_{Q}\equiv {\bf 0}$ but there exists $v\in l\cap Q$ with $w_{1}(v)=1$.
\item Connected components $R$ of $U$ such that $w_{1}|_{R}\equiv {\bf 0}$, $w_{2}|_{R}\neq {\bf 0}$.
\item Connected components $S$ of $U$ such that $w_{1}|_{S}\equiv {\bf 0}$, $w_{2}|_{S}\equiv {\bf 0}$.
\end{enumerate}

Since $H$ was connected, all such components contain at least one vertex from $l$. Moreover, we can assume that there exist no components of type (iv), since we can eliminate them from consideration as they do not affect the overall result. Also,  the induction step is trivial if $U$ has no components of type (iii). So assume that $U$ contains some component of type (iii).

\begin{itemize}
 \item {\bf Case 1: $w_{1}|_{l}\neq {\bf 0}$ and there exists $z\in l$ such that either $w_{1}(z)=0$ and $z$ belongs to a component of type (i) or $w_{2}(z)=0$ and $z$ belongs to a component of type (iii)}.

We update the state using the following sequence of steps.
\begin{itemize}
\item{1:} Schedule edge $l$. This is possible since the current configuration contains, by the hypothesis, at least a node $v\in l$ with $w_{1}(v)=1$. In turn, this changes the states of nodes in $l$ that belong to components of type (iii) to one.
\item{2:} Schedule components of type (iii), moving the states of their nodes from $w_{1}^{(l)}$ to $w_{2}^{(l)}$. This is possible since systems $U(w_{1},w_{2})$ and $U(w_{1}^{(l)},w_{2}^{(l)})$ are equivalent (thus satisfiable), and state $w_{1}^{(l)}$ contains a nonzero value on every connected component of type (iii).
\item{3:} Schedule edge $l$ again. This is possible since the state of node $z$ is 1. This makes the states of nodes $t$ in $l$ that belong to components of type (iii) take value $w_{2}(t)$. Also, this changes the values of other nodes $t\in l$ to $w_{1}(t)$ again.
\item{4:} Finally, schedule components of type (i) and (ii), changing the values of their nodes from $w_{1}$ to $w_{2}$. This is done componentwise, using the induction hypothesis, and the nonzero values of $w_{1}$ in each such component.
\end{itemize}

\item {\bf Case 2: $w_{1}|_{l}\neq {\bf 0}$, Case 1 doesn't apply and there exists $P$ a component of type (i) or (ii) with $w_{2}(z)=1$ for some $z\in l\cap P$}.

Since Case 1 does not apply, we can identify the values of $w_{1}$ on every vertex $x\in l$: $w_{1}(x)=1$ if $x$ belongs to a connected component of type (i) or (ii), $w_{1}(x)=0$ otherwise. Also, if $x\in l$ belongs to a configuration of type (iii), $w_{2}(x)=1$.

Consider an edge $l_{2}$ such that $l_{2}\cap l\neq \emptyset$ and $l_{2}$ belongs to a component of type (iii). $l_{2}$ exists (i.e. $Q$ does not consists of an isolated vertex in $U$) otherwise the component $Q$ would yield an unsatisfiable equation $0=1$ in $U(w_{1},w_{2})$.

\begin{itemize}
\item{1:} Schedule components of type (i) and (ii), changing the values of their nodes from $w_{1}$ to $w_{2}$.
\item{2:} Schedule edge $l$. This is possible since the current configuration contains, by the hypothesis, at least a node $z\in l$ with $w_{2}(z)=1$. In turn, this turns the states of nodes in $l$ that belong to components of type (iii) to one.
\item{3:} Schedule all components $P$ of type (iii) except $Q$, moving the states of their nodes from $w_{1}^{(l)}$ to $w_{2}^{(l)}$, restricted to the scheduled components. This is possible since systems $U(w_{1},w_{2})$ and $U(w_{1}^{(l)},w_{2}^{(l)})$ are equivalent.
\item{4:} Schedule component $Q$, moving its state from
$w_{1}^{(l)}|_{Q}$ to $(w_{2}^{(l)})|_{Q}$. This is
possible since system $U(w_{1}^{(l)},w_{2}^{(l)})$ is solvable.

Note that, since Case 1 doesn't apply, $w_{2}^{(l)}(x)=0$ for all $x\in Q\cap l$.
\item{5:} Let $r$ be a node of $Q$ with $w_{2}(r)=1$, closest to some vertex of edge $l$ in distance. Consider the path $D$ from $r$ to a node in $l$. All edges other than the one containing $r$ have only zero values. Schedule edges on path $D$ from $r$ towards $l$, one by one (Figure 3(a)). This ``propagates'' the value 1 towards the vertex in $l$. Note that all nodes on intermediate edges had initially zero values, so they have intermediate nodes on the path
retail value 1. In particular, last edge has a node in $Q\setminus \{l\}$ whose value is one.

 Also, this action turns the values of nodes in $l\cap Q$ to one.
\item{6:} Schedule edge $l$ again. This restores the correct value of all nodes in $l\setminus Q$. It also turns nodes in $l\setminus Q$ to zero.
\item{7:} ``Undo'' scheduling nodes on $D$ from $l$ to $r$ (Figure 3(b)). This is possible since, as each edge has $\geq 3$ vertices, last edge had a node whose value is one, and ``we propagate this value'' towards $r$. Since edges of $D$ were scheduled twice, every label is correct.
\end{itemize}

\begin{figure}[ht]
\hfill
  \begin{minipage}[t]{.45\textwidth}
\begin{center}
\epsfig{file=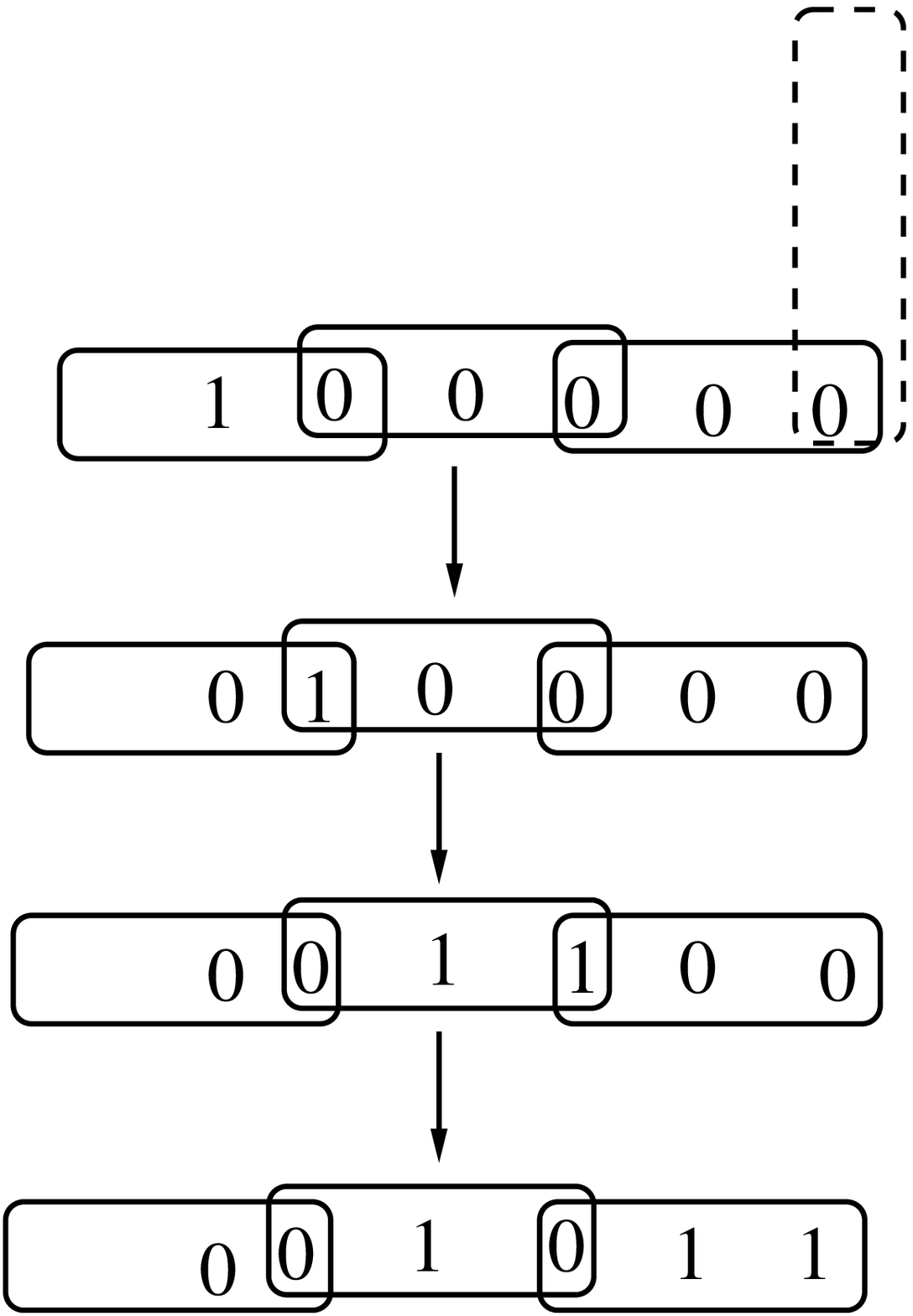,width=6cm}
\label{forward}
\end{center}
  \end{minipage}
  \hfill
  \begin{minipage}[t]{.45\textwidth}
    \begin{center}
\epsfig{file=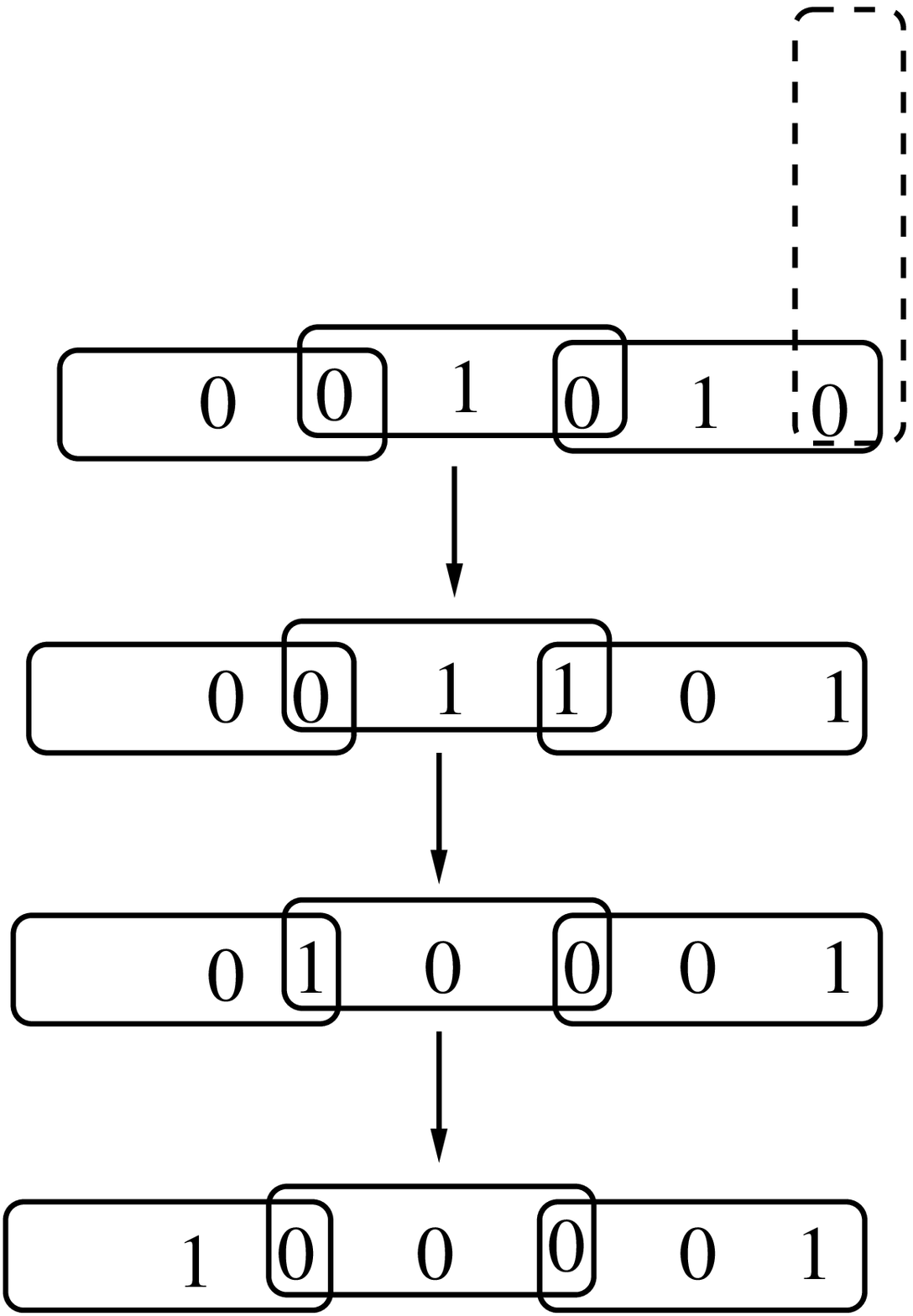,width=6cm}
\label{backward}
\end{center}
  \end{minipage}
\caption{(a). Forward propagation of ones. (b). Backward propagation. In both cases edge $l$ is shaded.}
\end{figure}
\item {\bf Case 3: $w_{1}|_{l}\neq {\bf 0}$ and Cases 1 and 2 do not apply.}

Since Cases 1  and 2 do not apply, we can identify the values of $w_{1},w_{2}$ on every vertex $x\in l$: $w_{1}(x)=1, w_{2}(x)=0$ if $x$ belongs to a connected component of type (i) or (ii), $w_{1}(x)=0$ otherwise. Also, if $x\in l$ belongs to a configuration of type (iii), $w_{1}(x)=0,w_{2}(x)=1$.

The construction is almost identical to that of Case 3. The only difference is that Step 1 is now executed at the very end of the process. This is possible, since before executing this step all vertices $x\in l$ belonging to components of type (i) or (ii) have value 1 (edge $l$ being scheduled twice was the only action affecting their value).

\item {\bf Case 4: $w_{1}|_{l}\equiv {\bf 0}$}. We reduce this case to one of the previous three cases as follows: since $w_{1}$ is not equal to ${\bf 0}$, there exists a vertex $v$ reachable from $l$ in $H$ with $w_{1}(v)=1$. Choose such a vertex $v$ at minimal distance from $l$ . Let this path be $D$. First, use the forward propagation trick on $D$ to change the state $w_{1}$ to a state $w_{3}$ with $w_{3}|_{l}\neq {\bf 0}$. System $H(w_{3},w_{2})=0$ has a solution with $z_{l}=0$, since system $H(w_{1},w_{2})=0$ has one. We then apply one of cases one to three, the one that works, to the former system.
\end{itemize}

\item {\bf For all edges $l$, all solutions of system $H(w_{1},w_{2})=0$ have $z_{l}=1$}.

If there exists $l$ with $w_{1}|_{l}\neq {\bf 0}$ and
$w_{1}^{(l)}\neq {\bf 0}$ then system $H(w_{1}^{(l)},w_{2})=0$ has a solution with $z_{l}=0$ (any solution of $H(w_{1},w_{2})=0$ with the value of $z_{l}$ flipped to zero). We then apply one of the previous cases.

Otherwise for all $l$ with $w_{1}|_{l}\neq {\bf 0}$,
$w_{1}^{(l)}\equiv {\bf 0}$. Therefore $H$ has a single such edge,
and the result immediately follows.

\end{enumerate}
\end{enumerate}
\endproof

\begin{lemma} \label{rec-reach} Let $H$ be a connected uniform hypergraph, and let $w_{1},w_{2},w_{3}$ be configurations on $H$.
If $w_{2}$ is reachable from $w_{1}$ and $w_{3}\neq {\bf 0}$ is reachable from $w_{1}$ in one step then $w_{2}$ is reachable from $w_{3}$.

Consequently one of the following two alternatives hold:
\begin{itemize}
 \item ${\bf 0}$ is reachable from $w_{1}$ and is the only recurrent state for the process started at $w_{1}$.
\item ${\bf 0}$ is not reachable from $w_{1}$ and $w_{2}$ is a recurrent state for the nondeterministic hyperedge switching process on $H$ started at $w_{1}$.
\end{itemize}
\end{lemma}
\beginproof
 If system $H(w_{1},w_{2})$ has a solution $(z_{e})$ and $l$ is the edge flipped when going from $w_{1}$ to $w_{3}$, let
\begin{equation}
 \overline{z_{e}}=\left\{\begin{array}{cc}
               z_{e},&\mbox { if }e\neq l, \\
           1+z_{l},&\mbox{ otherwise.}\\
              \end{array}
       \right.
\end{equation}
It is easy to see that $(\overline{z_{e}})$ is a solution to system $H(w_{3},w_{2})$. Since $w_{2}$ is special, the restriction in Lemma~\ref{onlyif} is satisfied for the pair $(w_{3},w_{2})$.
\endproof

Applying Lemma~\ref{onlyif} and Lemma~\ref{rec-reach} we obtain an easy algorithm for recurrence: make two reachability tests using Lemma~\ref{onlyif}.
\endproof

\section{Time to social balance on the triadic cycle}

Theorem~\ref{recurrent} motivates the study of the following quantity:

\begin{definition}
 For a triadic simplicial complex $G$, denote by $\tau_{SB}(G)$ the maximum, over all initial states $s$, of
$E[T_{s}(G)]$, where $T_{s}(G)$ is the expected time for the system, starting from initial state $s$ to enter a state with
zero (one) imbalanced triangles (according to the result in
Theorem~\ref{recurrent}). $\tau_{SB}(G)$ will be called {\em the time to social balance} of graph $G$.
\end{definition}

We first investigate the convergence time of random triad dynamics on the triadic cycle $T_{n}$. For this class of graphs, Theorem~\ref{recurrent} states that all recurrent states are socially balanced. We were motivated in our particular choice of graph $G$ by our paper \cite{ipd:colearning}, which investigated a similar dynamics. The dynamics in \cite{ipd:colearning} can also be described using the particle analogy, and is specified by particle creation/annihilation rules $(1,0)\goesto (1,1)$, $(0,1)\goesto (1,1)$, $(1,1)\goesto (0,0)$. In contrast, the annihilating random walk corresponding to the dynamics in this paper is specified by a very similar set of rules $(0,1)\goesto(1,0)$, $(1,0)\goesto (0,1)$, $(1,1)\goesto (0,0)$.

The next result shows that the convergence time to a socially balanced state is similar to the analog result in \cite{ipd:colearning}. The proof is simpler, though:

\begin{theorem}
Let $p_{-}=\min\{p,1/2\}$.
 The time to social balance on the triadic cycle $T_{n}$ is $O(\frac{n}{p^{-}})$.
\end{theorem}

\beginproof

The result follows from the following:

\begin{lemma}\label{dec-time}
 Function $P$ decreases at each step with probability
at least $p_{-}$.
\end{lemma}
\beginproof

\begin{figure}[h]
  \hfill
  \begin{minipage}[t]{.45\textwidth}
    \begin{center}
      \epsfig{file=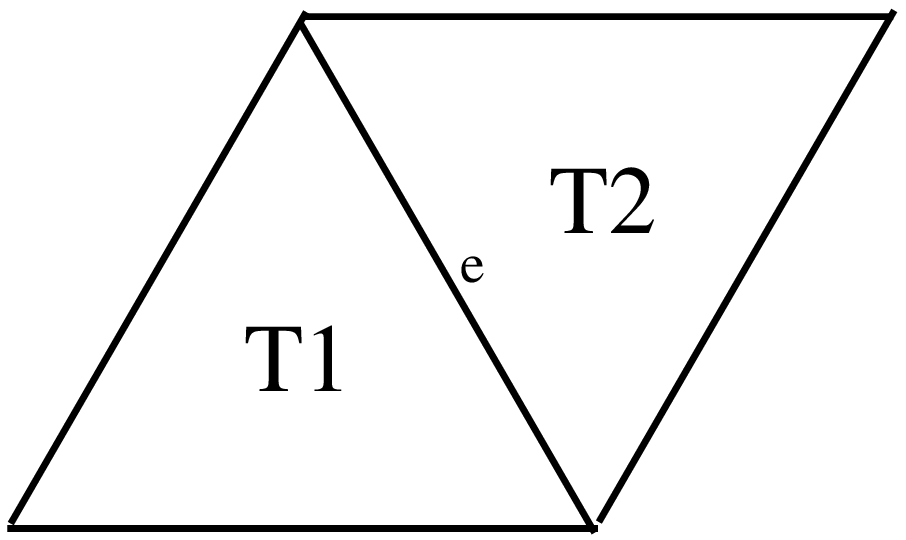, scale=0.45}
      \caption{The two triangles in the proof of Lemma~\ref{dec}.}
\label{two-triangles}
    \end{center}
  \end{minipage}
  \hfill
  \begin{minipage}[t]{.45\textwidth}
    \begin{center}
      \epsfig{file=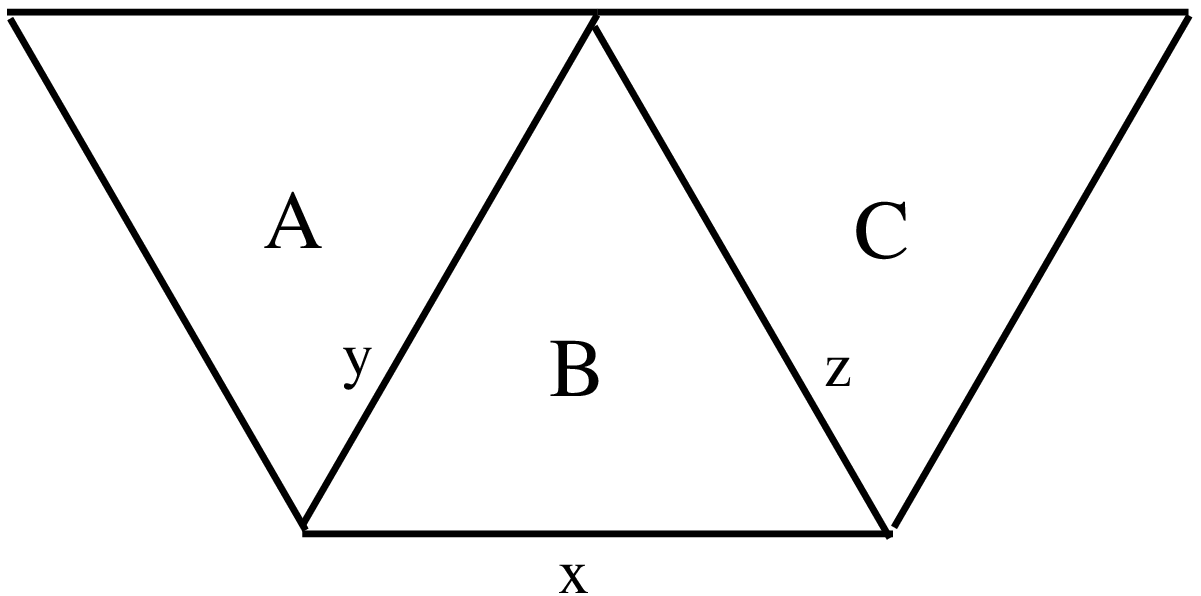, scale=0.45}
      \caption{The three triangles in the proof of Lemma~\ref{dec-time}.}
     \label{three-triangles}
    \end{center}
  \end{minipage}
  \hfill
\end{figure}

Consider three adjacent triangles in the graph $TC_{n}$, as displayed in Figure~\ref{three-triangles}. Suppose that triangle $B$ is unbalanced and is the one chosen by the dynamics. There are two cases:

\begin{enumerate}
 \item The dynamics changes the value of edge $x$ (the one that is not part of any other triangle). In this case no triangle but $B$ is affected, and the number of unbalanced triangles goes down by one.
\item The dynamics changes the value of one of the edges $y,z$.
In this case triangle $B$ becomes balanced. The triangle that shares
the chosen edge with $B$ becomes unbalanced if it was previously
balanced, and balanced if it was unbalanced as well. All in all, the
number of unbalanced triangles does not increase.
\end{enumerate}

\endproof

We now apply the following result (\cite{ragh-mot-randalgs} Theorem 1.3 pp. 15):

\begin{lemma}
 Let $g:{\bf R}_{+}\goesto {\bf R}_{+}$ be a monotone nondecreasing function. Consider a particle whose position changes at integer moments and is always an integer. If the particle is at position $m>1$, it proceeds at the next step to position $m-X$, where $X$ is a random variable ranging over the integers $1, \ldots, m-1$. All we know about $X$ is that $E[X]\geq g(m)$ and that $X$ is chosen independently of the past.

Let $T$ be the random variable denoting the number of steps in which the particle reaches position 1. Then $
 E[T]\leq \int_{1}^{n} \frac{dx}{g(x)}$.

\end{lemma}

Applying this result to the potential function counting the number
of unbalanced triangles in graph $T_{n}$, and employing
Lemma~\ref{dec-time} we infer that $\tau_{SB}(T_{n})= O(
\frac{n}{p^{-}})$.

\endproof

\section{The case $p=1/3$ on a 2-regular triadic simplex}

Another case where the dynamics is easy to analyze is that when $p =
1/3$ and graph $G$ is a 2-regular triadic simplex. In this case,
according to Theorem~\ref{recurrent}, all recurrent states are
either socially balanced (fixpoints), or all of them contain exactly
one imbalanced triangle. Radicchi et al. \cite{triad-lattice}
experimentally study the the convergence time on the triangular
lattice and obtain the estimate $\theta(N^{\alpha})$, with $\alpha
\sim 2.24$.

We can estimate the convergence time as follows:

\begin{theorem} \label{convergence}
 For any 2-regular triadic simplex $G$ such that $T_{3}(G)$ is connected there exists $C>0$ such that the time to social balance on $G$ satisfies
\begin{equation}
\tau_{M}(T_{3}(G))\leq \tau_{SB}(G)\leq T_{CRW}(T_{3}(G))\leq C
\cdot n^{min\{3,2+\log_{2}(\tau_{C}(T_{3}(G))\}},
\end{equation}  where $\tau_{C}(T_{3}(G))$ is the Cheeger constant of the triadic dual
 $T_{3}(G)$ and $n$ is the number of vertices of $T_{3}(G)$ (i.e.
 the number of {\em triangles} of graph $G$).
\end{theorem}

Before going into the proof, note that an open problem in \cite{aldous-fill-book} (Chapter 14, Open problem 13), combined with Proposition 5 (same chapter) would imply that the lower and upper bounds in Theorem~\ref{convergence} stated in terms of graph $T_{3}(G)$ have the same order of magnitude.

\beginproof

Since $G$ is 2-regular, $T_{3}(G)$ is a 3-regular {\em graph}, has no loops, and the clique switching process reduces to an annihilating random walk.

The annihilation time of an ARW is stochastically dominated by the coalescence time of a CRW on the same graph. This is an easy example of a coupling \cite{aldous-fill-book}:

\begin{lemma}
 We can couple the annihilating and coalescing random walk such that $T_{ARW}\leq T_{CRW}$. Consequently
\[
 T_{ARW}(G)\leq T_{CRW}(G).
\]

\end{lemma}
\beginproof

The coupling uses a well known idea (see e.g. \cite{griffeath-ips}). We first describe the process intuitively.
Consider a coalescing random walk. Declare particles that coalesce to be ``ghost'' particles (as opposed to the ``live'', not yet coalesced ones). If a set of ghost particle  meet a real particle they stick to it, adopting the (randomly chosen) trajectory of the live particle. When two live particle meet they become ghosts. The coupling is obtained by seeing the ARW as the CRW restricted to live particles.

Formally, define $(\xi^{x}_{t})_{x\in V,t\geq 0}$ to be a stochastic process such that
\begin{itemize}
 \item $\xi^{x}_{0}=x$ for all $x\in V$.
\item $\xi^{x}_{t}=\xi^{y}_{t}$ implies that $\xi^{x}_{t+s}=\xi^{y}_{t+s}$ for all $s \geq 0$.
\item For all $x\in V$, $(\xi^{x}_{t})_{t\geq 0}$ is a random walk on $G$.
\end{itemize}

Define the annihilating system $\eta_{t}\subseteq V$,
\[
 \eta_{t}=\{y|\mbox{ } |\{x\in V,\xi^{x}_{t}=y\}|\mbox{ is odd}\}.
\]

Assume that initially the number of particle was even. It is easy to see that  when all particles have coalesced then all particles are ghost particles. The reason is simple:
\begin{itemize}
 \item Any cluster of particle contains at most one live particle.
\item The only way for a live particle to become a ghost is to meet another live particle.
\item The number of live particle only goes down by two at a time.
\end{itemize}
\endproof

For the upper bound we apply a result due to Aldous and Fill (Proposition 9, Section  3.14, Chapter 14 in \cite{aldous-fill-book}), generalizing an earlier result of Donnelly and Welsh \cite{donnelly-welsh-interacting}\footnote{The result of Aldous and Fill is stated in \cite{aldous-fill-book} for continuous time, but can be easily translated to discrete time, at the expense of an additional linear factor}  and infer the theorem. We use the fact that the lattice graph is 3-edge-connected and 3-regular.

For the lower bound we consider all initial states consisting of exactly two imbalanced triangles, and use the connection with annihilating random walks.

\endproof

\section{Application to 3-XOR SAT}
Our problem also displays an unexpected connection with a problem in
the area of satisfiability solving: {\em random 3-XOR SAT}. The
satisfiability of random instances of this problem has been
investigated in both Statistical Physics \cite{zecchina:kxorsat} and
Theoretical Computer Science \cite{dubois-mandler-3xor}.

\begin{definition}
 An instance $F$ of the 3-XOR SAT problem is specified by a list of $m$ {\em equations} on $n$ variables, $x_{i_1}\oplus x_{i_2}\oplus x_{i_3}= b_{i}$, for some $b_{i}\in \{0,1\}$ and $i_{1},i_{2},i_{3}\in 1\ldots n$. $F$ is {\em satisfiable} if there exists an assignment $A$ of variables in $F$ that makes every equation evaluate to true.
\end{definition}

Our results will not involve random instances, but satisfiable
instances. Moreover, instances we will work with will be {\em
reduced}, i.e. they satisfy the following two conditions: no
variable appears in a single clause and no two clauses share two
variables. This last assumption does not particularly constrain the
class of instances we want to solve. Indeed, it is easy to see that
any formula can be transformed to a reduced one: We simply eliminate
clauses involving pure variables (since they can always be satisfied
by setting the pure variable the right way). Also, if $C_1$ and
$C_2$ share two variables, say $x$ and $y$, let $z$ and $t$ the
remaining variables. The conjunction of $C_{1}$ and $C_{2}$ entails
a constraint of the type $z\oplus t = \lambda$, for some $\lambda
\in \{0,1\}$. We can thus eliminate one of the variables $z$, $t$
(by replacing it with $t\oplus \lambda$). This eliminates one of the
clauses $C_1$ and $C_2$ from the formula.

We will analyze the RandomWalk algorithm displayed in
Figure~\ref{alg}. We will assume that the input to the RandomWalk
algorithm is a reduced formula with the following additional
properties:

\begin{definition}
 A 3-XOR formula is {\em connected} if one cannot partition $F$ into
 two variable-disjoint formulas. $F$ is {\em $k$-connected} if one can delete
 up to $k-1$ variables (and the clauses of $F$ where these variables appear)
 without disconnecting the formula. $F$ is {\em 2-regular} if every
 variable appears exactly in two clauses.
\end{definition}

The displayed connection with Coalescing Random Walks enables us to
prove the following result:

\begin{figure}
\dahntab{
\=\ \ \ \ \=\ \ \ \ \=\ \ \ \ \= \ \ \ \ \=\ \ \ \ \= \\
\hspace{5mm} {\bf Algorithm RandomWalkSat($\Phi$):} \\
\\
Start with an arbitrary assignment $U$. \\
while (there exists some unsatisfied clause)\\
\> \> pick a random unsatisfied clause $C$ \\
\> \> change the value of a random variable of $C$ in $U$\\
\keyw{return} assignment $U$. \\
}

\caption{The RandomWalkSat algorithm}\label{alg}

\end{figure}

\begin{theorem} \label{time-rw}
 If reduced formula $\Phi$ is satisfiable, has $m$ equations, $n$ variables, is $2$-regular  and $s$-connected then the expected time until RandomWalk finds a solution satisfies:
\[
 E [T_{RW}] \leq \min\{m^{3}/2s, 2\log 2\cdot \tau_{C}(T_{3}(\Phi))\cdot m^2\}.
\]
\end{theorem}

\beginproof
 Consider the hypergraph s$S(\Phi)$ associated to the reduced formula and its
 triadic dual $T_{\Phi}$. Since $\Phi$ is 2-regular it follows that
 $T_{\Phi}$ is actually a graph, and the dynamics of The RandomWalk algorithm
 on $\Phi$ can be interpreted as an annihilating random walk on $T_{\Phi}$. We then
 apply the (discrete time version of) results of Donnelly and Welsh, as extended by
Aldous and Fill in  Chapter 14 in \cite{aldous-fill-book}.
\endproof

\section{Conclusions}

Our paper raises the interesting prospect that the convergence time of local search algorithms, such as the RandomWalk algorithm might be analyzed with ideas from the interacting particle systems literature. We believe that this would be  especially interesting for ``message-passing'' algorithms
such as the belief and {\em survey propagation} algorithms
\cite{surveyprop-rsa2005}.

Clearly, we would like to see progress in analyzing the dynamics of the random hyperedge switching process on hypergraphs. In particular  studying the mixing time of the random hypergraph switching process is a  very interesting problem for further study. Another possible direction, motivated by the analogy with the dynamics studied in \cite{ipd:colearning}, is to further study the time to social balance using ideas similar to those in \cite{mossel-ipd}.

Finally, an interesting issue it to compute the Cheeger constant of the triangular lattice, as well as other planar regular lattices \cite{grunbaum-shephard}, and comparing the results with those in \cite{triad-lattice}.

\section*{Acknowledgments}
I thank  Cornel Izba\c{s}a and Cosmin Bonchi\c{s} for useful discussions.

This work has been supported by a Marie Curie International
Reintegration Grant within the 6th European Community Framework
Program and by the Romanian CNCSIS under a PN II/Parteneriate
grant.
\bibliographystyle{fundam}
\bibliography{C:/bib/bibtheory}
\end{document}